\documentclass[apm,reprint,amsmath,amssymb]{revtex4-1}

\usepackage{graphicx}
\usepackage{bm}

\begin{document}

\title{N-Type Oxide Thermoelectrics Via Visual Search Strategies}

\author{Guangzong Xing$^1$}
\author{Jifeng Sun$^2$}
\author{Khuong P. Ong$^3$}
\author{Xiaofeng Fan$^1$}
\author{Weitao Zheng$^1$}
\author{David J. Singh$^2$}

\affiliation{
$^1$College of Materials Science and Engineering, Jilin University, 130012, Changchun, China \\
$^2$Department of Physics and Astronomy, University of Missouri, Columbia, Missouri, 65211-7010, USA \\
$^3$Institute of High Performance Computing, Agency for Science Technology and Research, 1 Fusionopolis Way, 16-16 Connexis, 138632 Singapore \\
}

\date{\today}

\begin{abstract}
We discuss and present search strategies for finding new thermoelectric
compositions based on first principles
electronic structure and transport calculations. We illustrate them by
application to a
search for potential n-type oxide thermoelectric materials.
This includes a screen based on visualization of electronic energy
isosurfaces.
We report compounds that show potential as thermoelectric materials
along with detailed properties, including SrTiO$_3$, which is a known
thermoelectric, and appropriately doped KNbO$_3$ and rutile TiO$_2$.
\end{abstract}


\maketitle


Thermoelectrics are an energy conversion technology that is used in
spacecraft power generation, portable coolers and other areas.
\cite{wood,riffat,yang,disalvo}
The direct thermal to electrical energy conversion provided by
this technology and its scalability to low size and power suggests
numerous potential applications.
\cite{kraemer,kumar}
However, this is limited by the low efficiency of existing devices,
in turn limited by the performance of known thermoelectric
materials.
The maximum efficiency is governed by a dimensionless, but
temperature dependent figure of merit, $ZT$=$\sigma S^2T/\kappa$,
where $T$ is absolute temperature, $\sigma$ is electrical conductivity,
$S$ is thermopower (also known as the Seebeck coefficient) and
$\kappa$ is the thermal conductivity, usually written as
a sum of electronic and lattice contributions,
$\kappa$=$\kappa_e$+$\kappa_l$, which is typically a good approximation.

Thermoelectric materials pose a particular challenge to materials 
genome approaches. This is because thermoelectric performance,
characterized by a figure of merit, $ZT$, is a contraindicated
property of matter. In particular, standard common models, such
as the isotropic parabolic band model, with normal assumptions
predict that high $ZT$ will not occur.
This means that very high performance thermoelectric materials
have unusual electronic features, a number of which have been proposed
and identified as important in various cases.
These include reduced dimensionality, either in artificial structures
\cite{hicks}
or due to details of band formation, and other complex electronic
structure characteristics.
\cite{kuroki,chen,usui,usui-sto,shirai,parker-2d,mecholsky,pei,bjerg,ohkubo,shi}

Most good thermoelectric materials,
such as (Bi,Sb)$_3$Te$_2$, PbTe,
Si-Ge, skutterudites and
Mg$_2$(Si,Sn) are narrow to moderate gap semiconductors.
Several papers reporting searches for new semiconductors with thermoelectric
potential have been reported.
\cite{madsen-1,curtarolo,wang,bhattacharya}
However, there are several thermoelectric materials that do not
fit in the category of moderate band gap semiconductors, including some
oxides.
Oxides can potentially offer a number
of advantages for application including perhaps low cost ceramic
materials processing. \cite{fergus,koumoto}
The high $ZT$ oxide thermoelectrics are
Na$_x$CoO$_2$ and closely related p-type layered Co and Rh
oxides, with figures of merit approaching $ZT$$\sim$1.
\cite{terasaki,hebert,miyazaki,koumoto1}
The best $n$-type oxide thermoelectrics are very different, and the best
$n$-type $ZT$ values barely exceed 0.5. This is found in
ZnO ceramics at very high temperature. \cite{ohtaki,jood,ong}
This is an oxide semiconductor with a substantial gap.

A particular challenge in finding new oxide thermoelectrics is that
$ZT$ depends strongly on doping level, among other material parameters,
and therefore assessing the potential of a given material requires detailed
doping studies. Here we demonstrate 
some simple measures that can be readily calculated from first principles
and used in searching for new oxide thermoelectrics. For this purpose,
we use a set of early transition metal and simple oxides, selected with
a view to having oxides that are at least potentially dopable
n-type, based on the criterion that they either are known to be
dopable or that they have the same transition element and similar coordination
to a dopable material. The list includes two known transparent
conductors, In$_2$O$_3$ and BaSnO$_3$, several niobates and titanates
as well as BaZrO$_3$ for comparison.
We focus on high temperature because the thermoelectric performance
generally increases with temperature for wide band gap materials, 
and so if there are any high $ZT$ n-type oxide thermoelectrics
in this class, they are likely to have their best performance at
high temperatures.
We note that several of these materials have complex heavy valence
band structures, which might suggest p-type thermoelectricity. However,
these compounds are all naturally n-type, and p-type doping is unlikely
to be feasible; we do not pursue this further.

The calculations were performed consistently for all compounds.
The first principles calculations used the general potential linearized
augmented planewave (LAPW) method, \cite{singh-book}
as implemented in the WIEN2k code, \cite{wien2k}
with well converged LAPW basis sets, including local orbitals for
semicore states. Experimental lattice parameters were
used, while atomic coordinates were determined by energy minimization,
subject to the experimental space group symmetry. This was done
with the generalized gradient approximation functional of Perdew,
Burke and Ernzerhof (PBE-GGA). \cite{pbe}
Electronic structures were then calculated using the modified
Becke-Johnson potential of Tran and Blaha (mBJ), \cite{mbj}
which improves band gaps of these materials relative to PBE.
\cite{mbj,koller,singh1,kim-mbj,singh2}
Transport coefficients were obtained from the electronic structures
using the constant scattering time approximation and the BoltzTraP
code. \cite{boltztrap}
The Seebeck coefficients presented are all ceramic averages, defined by
$S_{av}=(\sigma_{xx}S_{xx}+\sigma_{yy}S_{yy}+\sigma_{zz}S_{zz})/
(\sigma_{xx}+\sigma_{yy}+\sigma_{zz})$, where $\sigma$ is the
electrical conductivity.

In order to develop screens for thermoelectric oxides, it is 
useful to begin with a parabolic band model.
For a degenerate doped single parabolic band (SPB), effective mass, $m^*$,
the Seebeck coefficient at low temperature ($k_BT\ll E_F$)
in the constant scattering time approximation is given by

\begin{equation}
S=\frac{8\pi^2k_B^2m^*T}{3e\hbar^2}\left (\frac{\pi}{3n} \right )^{2/3}=
\frac{\pi^2k_B}{2e}\frac{k_BT}{E_F} ,
\label{eqn-e}
\end{equation}

\noindent 
where $E_F$ is measured from the band edge (note $e$ is negative).
Thus $S(T)$ increases linearly with effective mass and temperature,
and decreases as the 2/3 power of carrier concentration.
The expression in terms of $E_F$ holds more generally
for a degenerate doped
single parabolic band, regardless of the anisotropy of the effective mass
tensor and importantly, when written in terms
of the energy, $S$ is independent of $m^*$.
\cite{kolodziejczak,parker-agbise2}
The volumetric density of states is 
$N(E)=(V/2\pi^2)(2m/\hbar^2)^{3/2}E_F^{1/2}$.

The conductivity for a doped SPB under the same conditions is given
by $\sigma=ne^2\tau/m^*$, where $\tau$ is an effective inverse scattering
rate. For an anisotropic parabolic band
$1/m^*$ is replaced by an inverse effective mass tensor and the
conductivity tensor takes the anisotropy of this tensor.
Thus $\sigma S^2$, which is in the numerator of $ZT$ varies as
$m^*\tau/n^{1/3}$. $\tau$ generally
decreases with both $m^*$ and $n$. Thus this formula does not mean
that high effective mass by itself is sufficient to get high $ZT$.

\begin{figure}
 \includegraphics[width=\columnwidth,angle=0]{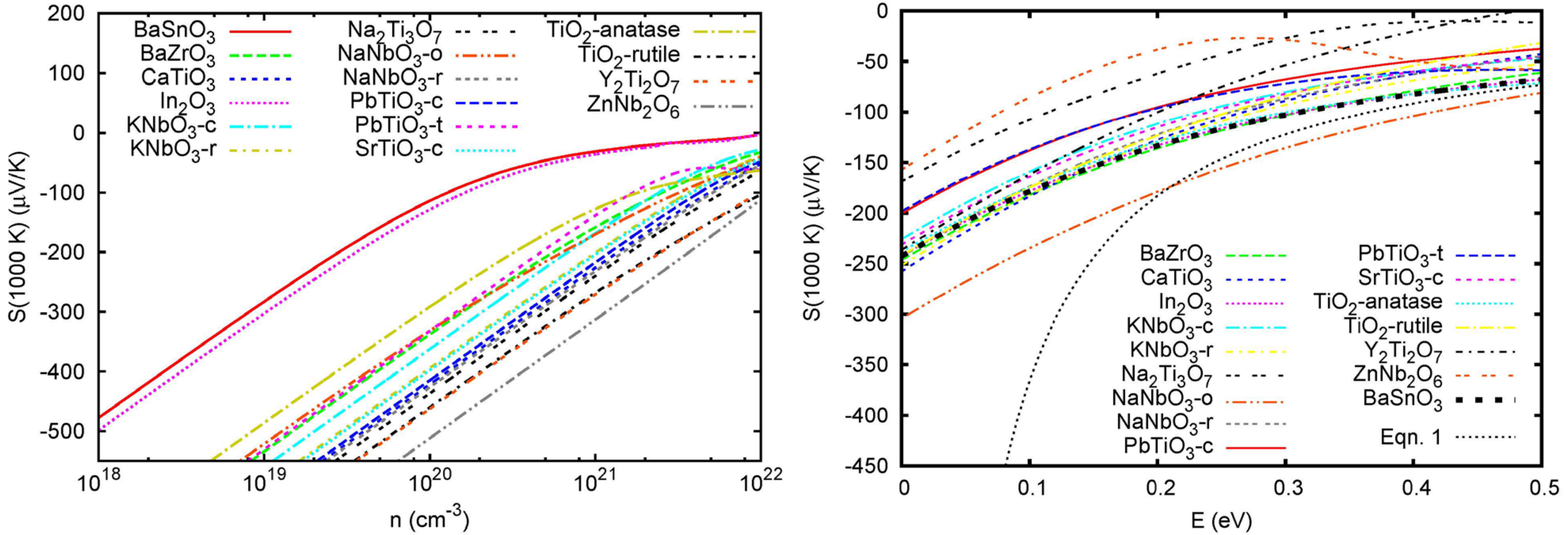}
\caption{$S$(1000 K) as a function of doping level on a log scale (left)
and as function of Fermi energy, $E_F$ (right). For non-cubic materials the
thermopower shown is the ceramic average, 
$S_{av}=(\sigma_{xx}S_{xx}+\sigma_{yy}S_{yy}+\sigma_{zz}S_{zz})/
(\sigma_{xx}+\sigma_{yy}+\sigma_{zz})$. In the right panel the limiting
formula, Eqn. 1 is also plotted, and a heavier dashed line is shown for
BaSnO$_3$, which has a single almost perfectly isotropic parabolic
band at the CBM, illustrating the behavior of a SPB.
}
\label{fig-see}
\end{figure}

These formulas imply that if one considers different semiconducting
compounds one should expect a wide range of doping dependent thermopowers.
This is seen in the left panel of
Fig. \ref{fig-see}, which shows the doping dependent $S$(1000 K)
for the compounds investigated here. On the other
hand, the various compounds appear more similar at fixed
energy. This is
seen in the right panel of the figure, which shows the $S$(1000 K)
as a function of Fermi energy, $E_F$ relative to the conduction
band minimum, and also in Table \ref{tab-compounds}.
The thermopower at fixed Fermi level and temperature is relatively
constant between the different phases, and in particular there is no
correlation with the density of states, $N(E_F)$. The outlier compounds
in right panel of Fig. \ref{fig-see} are Na$_2$Ti$_3$O$_7$ and ZnNb$_2$O$_6$,
for low $S$ and orthorhombic NaNbO$_3$ for high $S$. 
Orthorhombic NaNbO$_3$ is the polar
phase that exists at room temperature 
but becomes cubic, similar to KNbO$_3$ at high temperature.
\cite{wood-2,johnston}
The electronic structures of cubic NaNbO$_3$ and cubic KNbO$_3$ are very
similar.

\begin{table}
\caption{Properties of the oxides investigated, including
calculated band gap, $E_g$ in eV, and electronic properties 0.1 eV above the
CBM. These are the
density of states, $N$ in eV$^{-1}$ on a per atom basis,
ceramic average Seebeck at 1000 K, in $\mu$V/K,
direction averaged $\sigma/\tau$ in 10$^{18}$ $\Omega$m/s,
carrier concentration, n in
10$^{19}$ cm$^{-3}$ and conductivity anisotropy, $a=\sigma_{max}/\sigma_{min}$.
}
\label{tab-compounds}
\begin{tabular}{lcccccc}
\toprule
 & $E_g$~~ & ~~$N$~~ & ~~~S~~~ & ~$\sigma/\tau$~ & ~~$n$~~ & ~~$a$~ \\
\colrule
BaSnO$_3$ $Pm\bar{3}m$     & 2.8 & 0.0039 & -179 & ~3.7 & ~3.8 & 1 \\
BaZrO$_3$ $Pm\bar{3}m$     & 4.3 & 0.076  & -182 & 28.4 & 71.6 & 1 \\
In$_2$O$_3$ $Ia\bar{3}$    & 3.2 & 0.0047 & -178 & ~4.0 & ~4.8 & 1 \\
KNbO$_3$-c $Pm\bar{3}m$    & 2.4 & 0.147  & -159 & 51.4 & 113 & 1 \\
KNbO$_3$-r $R3m$           & 2.8 & 0.121  & -181 & 38.8 & 128 & 1.09 \\
NaNbO$_3$-o $Pmc2_1$       & 2.4 & 0.013  & -235 & 12.0 & 37.8 & 1.61 \\
NaNbO$_3$-r $R3c$          & 3.4 & 0.160  & -174 & 40.3 & 203 & 1.34 \\
ZnNb$_2$O$_6$ $Pbcn$       & 4.1 & 2.23   & ~-85 & 10.4 & 1344 & 3.67 \\
CaTiO$_3$  $Pnma$          & 3.2 & 0.146  & -183 & 34.6 & 170 & 1.11 \\
Na$_2$Ti$_3$O$_7$ $P2_1/m$ & 4.0 & 0.304  & -108 & 11.2 & 532 & 979 \\
PbTiO$_3$-c $Pm\bar{3}m$   & 2.0 & 0.262  & -138 & 62.4 & 261 & 1 \\
PbTiO$_3$-t $P4mm$         & 2.1 & 0.102  & -137 & 20.6 & 102 & 299 \\
SrTiO$_3$-c $Pm\bar{3}m$   & 2.7 & 0.158  & -164 & 45.8 & 166 & 1 \\
TiO$_2$ (ana.) $I4_1/amd$  & 2.9 & 0.034  & -174 & 11.0 & 46.7 & 10.7 \\
TiO$_2$ (rut.) $P4_2/mnm$  & 2.5 & 0.191  & -175 & 56.7 & 344 & 1.01 \\
Y$_2$Ti$_2$O$_7$ $Fd\bar{3}m$ & 3.6 & 0.102  & -161 & 19.2 & 419 & 1 \\
\botrule
\end{tabular}
\end{table}

The effective mass plays different roles in $S$ and $\sigma/\tau$, as is
evident from the different effects of anisotropy in $m^*$ on
these two quantities. Also, as mentioned,
most good thermoelectrics have band structures that differ markedly from
the single isotropic parabolic band. A computational screen
for potential thermoelectrics may then begin by looking for band structures
that deviate from the isotropic SPB in ways that favor thermoelectric
performance.
To understand what kinds of band structures these are, we write the
Boltzmann transport formulas for the conductivity and thermopower starting
with the transport function,
\begin{equation}
\sigma_{\alpha\beta}(E)=e^2  \int 
{\rm d}^3{\bf k} ~v_\alpha({\bf k})v_\beta({\bf k})\tau({\bf k})
\delta(E-E({\bf k})) ,
\end{equation}
\noindent where summation over bands is implied,
$E({\bf k})$ is the band energy and $v=\hbar^{-1}\nabla_{\bf k}E$ is the 
band velocity. Then the conductivity and Seebeck tensors are
\begin{equation}
{\bf \sigma}(T)= -\int \sigma(E) f'(T,E-E_F) dE
\label{eqn-sig}
\end{equation}
\noindent and
\begin{equation}
{\bf S}(T)=\frac{-{\bf \sigma}^{-1}(T)}{eT}
\int \sigma(E) (E-E_F) f'(T,E-E_F) dE
\label{eqn-s}
\end{equation}
\noindent where $f'$ is its energy derivative of the Fermi function
at temperature $T$.

\begin{figure}
 \includegraphics[width=\columnwidth,angle=0]{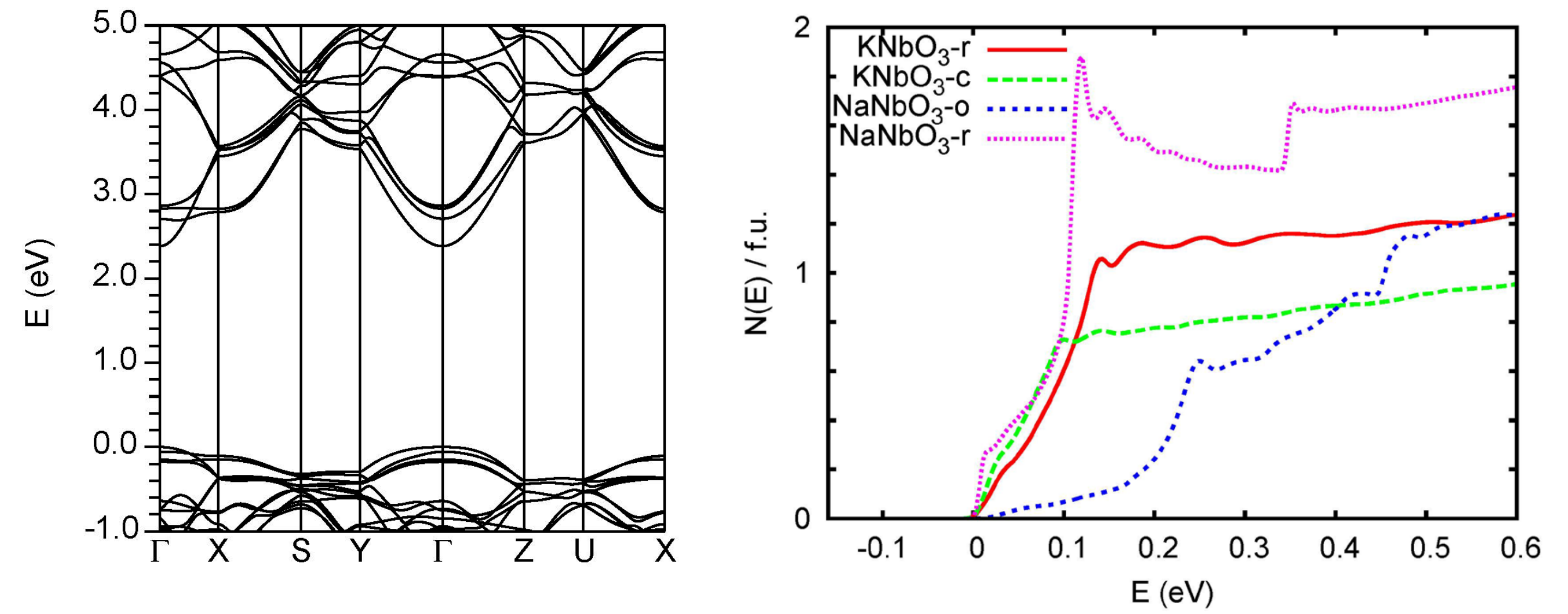}
\caption{Band structure (left) of orthorhombic NaNbO$_3$ and
DOS (right) for different structures of NaNbO$_3$ and KNbO$_3$ on a per
formula unit base, with the energy zero set to the CBM.}
\label{fig-niobate}
\end{figure}

Thus $\sigma$ and $S$ involve different integrals over the electronic
structure. There are two important aspects that we use here:
(1) At low temperature, Eqns. \ref{eqn-sig} and \ref{eqn-s} have different
integrals over the Fermi surface, i.e. the energy isosurface for
$E=E_F$. The integrals are such that regions of high $v_\alpha^2$
are highly weighted in $\sigma_{\alpha\alpha}$. For a closed isosurface
these are the parts where the span across the surface is small. From
this point of view, low dimensional and complex shapes are much superior
to simple shapes, such as the spherical isosurface of the isotropic SPB.
For example, a cylindrical isosurface is much more favorable than
a sphere enclosing the same volume (number of carriers) for conduction
across the cylinder. As discussed for PbTe and related thermoelectrics,
\cite{parker-2d,chen} isosurfaces with cylindrical sections in different
directions can yield isotropic properties that are far superior to those
that would be obtained in an SPB with the same carrier concentration.
More generally, complex, corrugated isosurface shapes lead to better
thermoelectric performance than the spheres that occur with an SPB.
\cite{kuroki,chen,usui,parker-2d,mecholsky,pei,ohkubo,shi}
(2) At finite $T$, the factor $E-E_F$ makes $S$ sensitive to the electronic
structure over a wider energy range than $\sigma$. As a result,
features in the electronic structure such as the onset of a heavy band
near $E_F$ can enhance $S$ while affecting $\sigma$ to a lesser extent.
Examples of thermoelectrics with a light band extremum followed by onset of
heavy band(s) include p-type filled skutterudites \cite{singh-skutt}
and n-type lanthanum telluride. \cite{may}
Polar orthorhombic NaNbO$_3$ is an example of such a case, and
this is the reason why it is the outlier with high $S$ in
Fig. \ref{fig-see}. Its band structure and density of states (DOS) in
comparison with some other niobates is shown in Fig. \ref{fig-niobate}.
The structure of the DOS is similar to p-type PbTe,
\cite{singh-pbte}
in contrast to the
sharp onsets characteristic of the low dimensional electronic features
of the cubic and rhombohedral perovskite niobates (see below).
Unfortunately, as mentioned, NaNbO$_3$ is not in the orthorhombic
polar phase at high temperature, and the particular characteristic
of a heavy band onset above a light mass CBM is a result of the
polar distortion. As such, while it is an interesting illustration
of this effect, we do not discuss this compound further.

\begin{figure}
 \includegraphics[width=\columnwidth,angle=0]{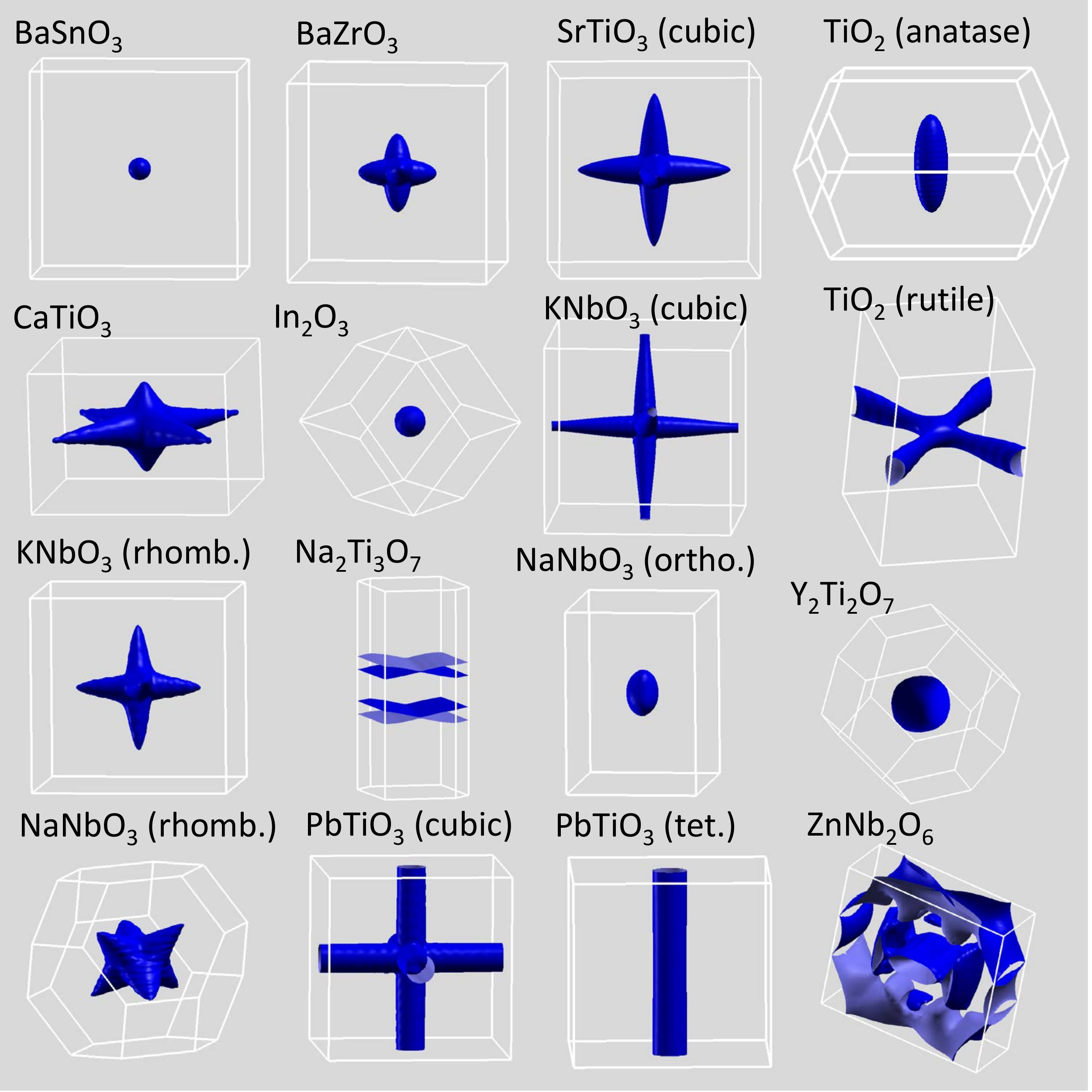}
\caption{Isoenergy surfaces 100 meV above the conduction band minimum
for candidate oxide materials. Several of the compounds also have smaller
isosurface sheets enclosed inside the sheet shown (see text).  }
\label{fig-fermi}
\end{figure}

Most high $ZT$ thermoelectrics have $S\sim$ 200 -- 300 $\mu$V/K at the
temperatures where they are used.
This is readily understood if one applies the Wiedemann-Franz
relation to the electronic part of the
the thermal conductivity to obtain
$ZT=\sigma S^2T/(\kappa_l+\kappa_e)=rS^2/L$,
where $r=\kappa_e/(\kappa_l+\kappa_e)\leq 1$ and $L$ is the Lorentz number.
The standard value of $L$ yields $ZT$=1 for $r$=1 and $S$=157 $\mu$V/K,
or for a typical $r\sim$0.5 with $S\sim$220 $\mu$V/K.
Considering the results shown in
Fig. \ref{fig-see}, this suggests a rough estimate of $E_F\sim$ 0.1 eV
for 1000 K. Isosurfaces 0.1 eV above the conduction band 
minimum (CBM) are shown in Fig. \ref{fig-fermi}. Examination of these
isosurfaces provides a very simple and fast screen since they are
based on the band structure alone.
For example, ZnNb$_2$O$_6$ shows a large complex isosurface.
This is characteristic of a metal, and this is why $S$ at fixed
$E_F$ is low. Moreover, having a large complex Fermi surface at $E_F$=0.1 eV
means that the conduction bands are extremely flat in this compound, as is 
also evident from the high $N(E_F)$.
Na$_2$Ti$_3$O$_7$ shows a near one dimensional metallic structure,
while the cylinder for tetragonal PbTiO$_3$ is a two dimensional electronic
feature. These materials will therefore be highly anisotropic, which is 
not favorable for making a ceramic thermoelectric. Anatase TiO$_2$
has an elongated ellipsoidal shape, which again leads to large conductivity
anisotropy. The transparent conductors, In$_2$O$_3$ and BaSnO$_3$ have small
simple isosurfaces, which according to the arguments above are unfavorable,
and this is also the case for pyrochlore Y$_2$Ti$_2$O$_7$.

The remaining
materials have complex isosurfaces anticipated to be
favorable for thermoelectric performance.
The $d^0$ cubic perovskites and
some distorted perovskites, i.e. rhombohedral KNbO$_3$
and NaNbO$_3$ and $Pnma$ CaTiO$_3$,
have these complex shapes because of the degeneracy
of the $t_{2g}$ orbitals in an octahedral crystal field, as was discussed
previously for SrTiO$_3$ and KTaO$_3$. \cite{usui-sto,shirai,fan}
Of the compounds investigated here, cubic KNbO$_3$, which is the high
temperature phase, shows the most perfect low dimensional character
in its band structure. This is seen in isosurfaces consisting of intersecting
nearly perfect cylinders for the lowest conduction band (note that 
the isosurfaces are a large section consisting of three cylinders joined
at the center around $\Gamma$ and two small sections inside formed by
the crossings of the three cylinders; in general cubic $d^0$ perovskites
have a large jack shaped section, as seen in SrTiO$_3$ with two small
section inside the center of the jack).
Interestingly, rutile TiO$_2$ also shows a low dimensional character,
favorable for thermoelectric performance,
with isosurfaces consisting of two intersecting cylinders
(note that two intersecting two dimensional, cylindrical sheets,
mean that the conductivity will be three dimensional, and in TiO$_2$
the calculated conductivity anisotropy is weak, as shown in
Table \ref{tab-compounds}).

\begin{figure}
 \includegraphics[width=\columnwidth,angle=0]{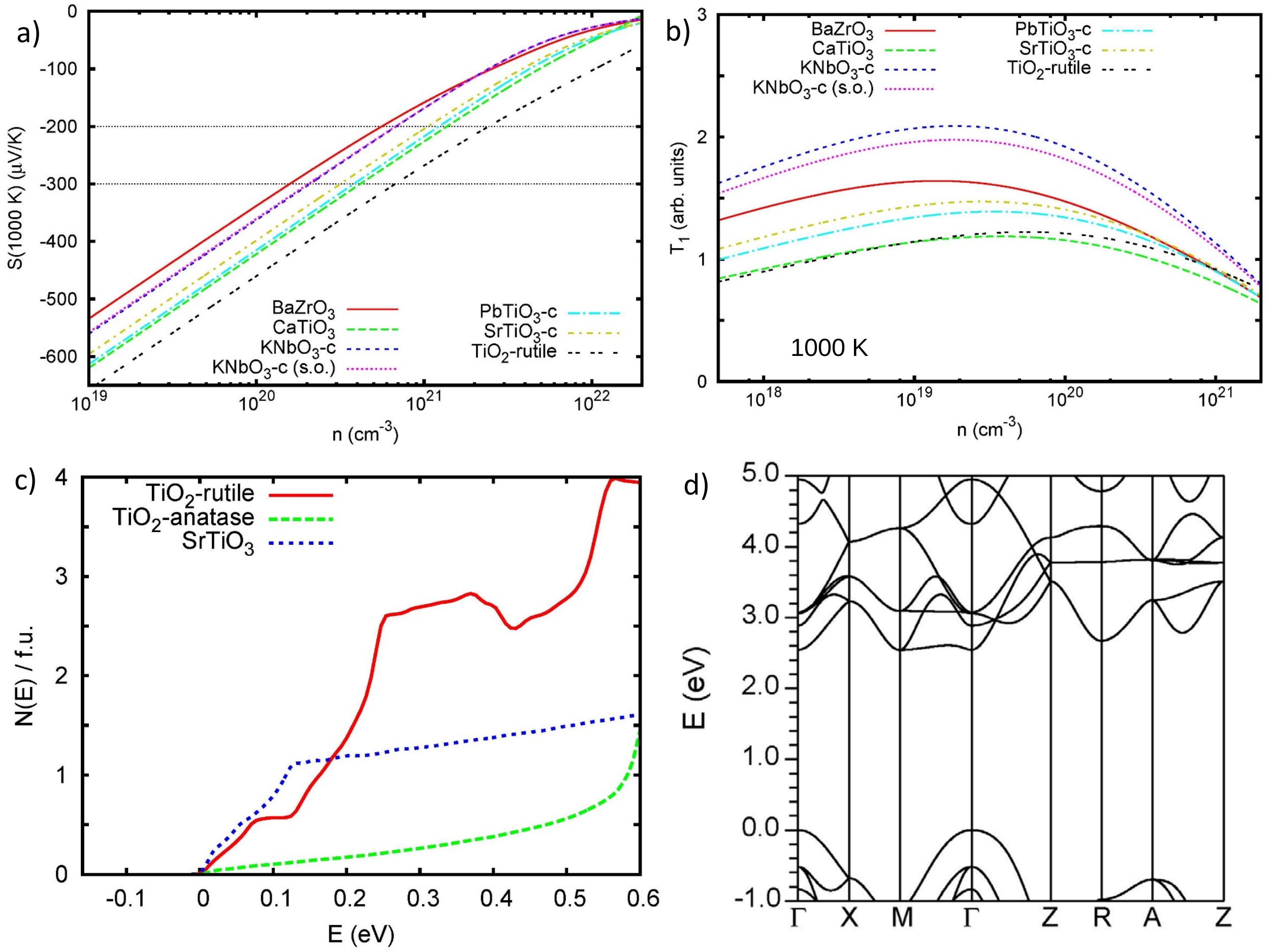}
\caption{(a) Doping dependent thermopower at 1000 K for the best identified
materials, (b) transport function $T_1$ (see text), (c) DOS for
TiO$_2$ and cubic SrTiO$_3$, (d) band structure of rutile TiO$_2$.}
\label{fig-out}
\end{figure}

The top panels of Fig. \ref{fig-out} show transport properties for
the compounds identified based on the above screen. Additionally,
in the case of KNbO$_3$, which contains the 4$d$ element Nb,
we show results of calculations including spin-orbit. As seen, these are
very similar to the scalar relativistic results.
The doping range of interest for developing these materials as thermoelectrics
at 1000 K ranges from $\sim$10$^{20}$ cm$^{-3}$ for BaZrO$_3$ and
cubic KNbO$_3$ to $\sim$2x10$^{21}$ for TiO$_2$.
This will shift to lower values for optimization at lower $T$, but as
mentioned for these wide gap materials the best performance is likely to
be close to the highest temperature at which the material can be used.
The transport function $T_1$ plotted in the top right panel
is $(\sigma/\tau)S^2/N^{2/3}$, where $N$ is the volumetric density of
states. The rational for using this particular function is that
it is readily computed from the band structure, that the 2/3 power
of the DOS cancels the mass dependence in the numerator for a SPB,
and that while scattering depends strongly on the details of a material,
the 2/3 power law dependence of the scattering rate on $N$ is a reasonable
choice for a degenerate doped semiconductor at high temperature.
This is clearly a crude measure of the potential electronic
behavior of a thermoelectric, since it makes no attempt to incorporate
detailed material specific scattering mechanisms. However, it does suggest that
KNbO$_3$ may be the best of the materials studied if it can be doped.
Also, interestingly, the non-perovskite, rutile TiO$_2$ is within a factor
of two of this ``best" material. Anatase on the other hand is not favorable.
The bottom left panel shows the DOS for rutile and anatase as compared
with SrTiO$_3$. As seen, SrTiO$_3$ and rutile have a similar
sharp 2D-like onset at the CBM, while anatase has a lower, more 3D shape
(for a 2D band, $N(E)$ has a step function shape, while for a 3D
parabolic band $N(E)\propto E^{1/2}$).
Bayerl and Kioupakis have reported theoretical investigation
by a somewhat different approach of different
TiO$_2$ polymorphs as potential n-type thermoelectrics and also find that
rutile is exceptional. \cite{bayerl}
The band structure of TiO$_2$ (bottom, right) has a double degenerate
bottom conduction band at $M$, with a very flat CBM along the $\Gamma$-$M$
directions. This underlies the favorable behavior.

From an experimental point of view, SrTiO$_3$ is known to be dopable
using Nb, and $ZT\sim 0.37$ has been reported at 1000 K with 20\% Nb,
\cite{ohta-sto,wang-sto}
This corresponds to a very high carrier concentration, $\sim$3x10$^{21}$,
assuming one electron per Nb. The bands at the CBM are Ti $d$ bands,
specifically from the $t_{2g}$ level, which suggests exploration of
alternate doping strategies involving the O or Sr sites.
SrTiO$_3$, like other perovskites near ferroelectricity, has a low thermal
conductivity, which is favorable for a thermoelectric.
\cite{ohta-sto,wang-sto,tachibana}
The thermal conductivity of NaNbO$_3$ is particularly low at $\sim$1 W/mK
at 300 K, while KNbO$_3$ has $\kappa\sim$10 W/mK at this temperature.
\cite{tachibana}
Meanwhile the thermal conductivity of single crystal
TiO$_2$ at 300 K is $\sim$8 W/mK and falls with temperature to below
5 W/mK at 1000 K.
\cite{thurber}
Furthermore it is known that TiO$_2$ can be doped
n-type by Nb similar to SrTiO$_3$. \cite{thurber}

BaZrO$_3$ is an insulating material, with a higher
band gap than SrTiO$_3$ and it is not known whether
it can be effectively doped. Cubic PbTiO$_3$ has a ferroelectric transition
to the tetragonal form at $\sim$800 K, accompanied by a large $\sim$6\%
lattice strain, which makes it unlikely that practical thermoelectric
modules could be made based on this material.
Furthermore, heavy n-type doping of PbTiO$_3$ may be particularly challenging
because of the presence of Pb$^{2+}$, which might lead to low energy
compensating defects associated with availability of the Pb$^{4+}$ valence
state.

The results suggest that in addition to SrTiO$_3$, KNbO$_3$,
and perhaps its alloys such as (K,Na)NbO$_3$ may be particularly
good candidate materials. We also find that rutile TiO$_2$ is a 
good candidate and is remarkable in that it is a binary, non-perovskite
based compound.

The key challenge in developing any of these oxides into high performance
thermoelectrics will be doping. As mentioned, in SrTiO$_3$ and TiO$_2$
doping can be done by Nb alloying on the Ti site. However, the CBM's
have Ti $d$ character, which means that other doping strategies should be
explored. Both YTiO$_3$ and LaTiO$_3$ form in perovskite structure and
so exploration of the solid solutions with SrTiO$_3$ would be one avenue.
Another would be doping on the O site, perhaps with F.

In the case of TiO$_2$, the possibilities other than Nb doping are more
limited, but again it may be possible to use the O site.
This is consistent with transport measurements on titania doped
by oxygen deficiency.
\cite{monika2}
TiOF can be formed
in a tetragonal rutile structure, \cite{chamberland}
suggesting that there could be a high solubility of F in rutile TiO$_2$.
One issue with TiO$_2$ is the tendency of the binary oxide to oxygen
deficiency when processed at high temperature under reducing conditions.
This O deficiency, if large,
can lead to ordered phases, Ti$_n$O$_{2n-1}$, known as Magneli phases.
\cite{andersson,andersson2,walsh}
These conducting phases are generally made
by reduction of TiO$_2$ at temperatures
above 1300 K. \cite{walsh}
While these phases have several desirable properties for thermoelectrics,
so far the thermopowers obtained are too low to obtain high $ZT$.
\cite{harada,portehault,monika,monika2}
It would be important to avoid the
formation of Magneli phases in assessing doped rutile TiO$_2$ as a
thermoelectric. F doping and avoidance of high temperature anneals in 
reducing environments may be useful for this.
Doping of KNbO$_3$, or (K,Na)NbO$_3$ is also expected to be a challenge,
although the required doping level is lower than for the other compounds
studied. Some possibilities could be divalent substitutions on the K site,
such as Ba or again F on the O site. In this regard, it is known that 
BaNbO$_3$ forms in a perovskite structure and is metallic.
\cite{casais}

In summary, we present simple screens that can be readily
applied in searching for
new oxide thermoelectrics and demonstrate them by application to a set of
oxides.
This includes a simple screen based on energy isosurface
structure. This provides a visual search strategy.
We find that KNbO$_3$ and rutile TiO$_2$ are interesting compounds
to investigate further as potential thermoelectrics in addition to the
known thermoelectric compound SrTiO$_3$.

Work at the University of Missouri
(JS, DJS) is supported by the U.S. Department of Energy,
Basic Energy Sciences through the S3TEC Energy Frontier Research
Center, award DE-SC0001299/DE-FG02-09ER46577.
KPO is grateful for support from the Institute of High Performance
Computing (IHPC) A*STAR.

%

\section*{Figure Captions}
\pagebreak

\end{document}